\documentclass[apj]{emulateapj}
\setkeys{Gin}{width=0.48\textwidth}
\usepackage{graphicx}

\newcommand{\hmpc}{\ifmmode{h^{-1}\,\hbox{Mpc}}\else{$h^{-1}$\thinspace Mpc}\fi}

\newcommand{\kms}{\ifmmode{\,\hbox{km\,s}^{-1}}\else {\rm\,km\,s$^{-1}$}\fi}
\newcommand{\msun}{{\rm\,M_\odot}}

\begin{document}
\title{Velocity Variations in the Phoenix-Hermus Star Stream} 
\shorttitle{Phoenix-Hermus Velocities}
\shortauthors{Carlberg \& Grillmair}
\author{R. G. Carlberg}
\affil{Department of Astronomy \& Astrophysics, University of Toronto, Toronto, ON M5S 3H4, Canada} 
\email{carlberg@astro.utoronto.ca }
\author{C. J. Grillmair }
\affil{Spitzer Science Center, 1200 E. California Blvd., Pasadena, CA 91125, USA}
\email{carl@ipac.caltech.edu}

\begin{abstract}
Measurements of velocity and density perturbations along stellar streams in
the Milky Way  provide a time integrated measure 
of dark matter substructure at  larger galactic radius than 
the complementary instantaneous inner halo strong lensing detections of dark matter sub-halos in distant galaxies.
An interesting case to consider is the proposed Phoenix-Hermus star stream, 
which is long, thin and on a nearly circular orbit, making it a particular good target to study for velocity variations
along its length. 
In the presence of
dark matter sub-halos the stream velocities are significantly perturbed in a manner  
that is readily understood with the impulse approximation. 
A set of simulations show that only sub-halos above a few $10^7 \msun$ lead to reasonably long lived observationally detectable velocity variations  of amplitude of order 1 \kms, with an average of about 
one visible hit per (two-armed) stream over the last 3 Gyr. 
An implication is that globular clusters themselves will  not have a visible impact on the stream.
Radial velocities have the benefit that they are completely insensitive to distance errors. 
 Distance errors scatter individual star velocities perpendicular and tangential to the mean orbit
 but their mean values remain unbiased. Calculations like these help build the quantitative case to acquire
  large, fairly deep, precision velocity samples of  stream stars.
 
\end{abstract}

\section{INTRODUCTION}
\nobreak

The Milky Way's dark matter halo has long been known to contain dark matter sub-halos around 
its dwarf galaxies
\citep{Aaronson:83,FaberLin:83}.  The numbers of the most massive dwarf galaxies are comparable to the predictions
of LCDM galactic halos \citep{Moore:99,Klypin:99},  but are increasingly below the predictions as the mass 
(or a reference velocity) becomes smaller. 
The ``missing" dark matter sub-halos could either be present, but with a sufficiently low gas or star content 
that they emit no detectable radiation, or, the dwarf galaxy proxy of the sub-halo census is complete, 
indicating either the sub-halos were formed and subsequently destroyed by astrophysical processes, 
or, the LCDM cosmological model is incorrect on small scales so that large numbers of low mass dark halos 
never formed or survived the merger into larger galactic halos. 
At very high redshifts measurements of the UV luminosity function find that the slope
of the luminosity function steepens to values much closer to what LCDM predicts \citep{Schenker:13,Bouwens:15}.
Although the high redshift observations do not directly probe the lower mass dark halos that are apparently missing
at low redshift, the much steeper luminosity function is at least suggestive that 
many lower mass dark matter halos were present at  earlier times, 
suggesting that they still may be present, but 
contain too little visible material to be readily detected.

The inner  30 kpc  of a dark halo, where most currently known star streams largely orbit,  is
 fairly hostile to dark matter sub-halos, simply because
of the relatively strong tidal field. 
Consequently, the
fraction of the mass in sub-halos declines from 5\%  at the virial radius,
 to about 1\% at 30 kpc and 0.1\% at 8 kpc
\citep{Aquarius}. 
Strong lensing
is  an attractive technique that directly ``illuminates" the instantaneous position and size
of those sub-halos that are near the critical line of a strong lensing system \citep{Vegetti:14,Hezaveh:16}.
The critical line is generally at a fairly small galactic radius, for instance, in the strong lens  system SDP.81 the critical line is at a projected radial distance of 7.7 kpc. At such radii the 
sub-halo population is small and has a large  fractional variance \citep{Ishiyama:09,Chen:11} for a dark matter potential alone. 
In addition, inside of 8 kpc the presence of dense gas and the stellar structure of a galaxy will  act to further erode sub-halos and add variance to their numbers \citep{Donghia:10}.
On the other hand, thin stellar streams are currently known in the region between about 15 and 30 kpc, where local gas and stars are  not a significant part of the gravitational field.  

Streams have both the drawback and advantage that they  
integrate the encounters of sub-halos over their last few orbits, and do not give an instantaneous snapshot of where any particular dark halo is to be found.
A sub-halo crosses a stream of stars so quickly that the orbital changes of the stars in the vicinity can be derived from 
the impact approximation. Over about an orbital period
a gap opens up in the stream \citep{Carlberg:12,Erkal:15,Erkal:15b,Sanders:16,Erkal:16}. 
With time, the gaps are blurred out as the differential angular momentum in the stream
stars cause stars to move  into the gap in configuration space, 
with the blurring being faster for relatively more eccentric orbits. 

The main purpose of this paper
is to provide quantitative estimates of what can be expected as velocity data are acquired for
the Phoenix-Hermus stream which is likely an especially good case for study. 
Besides improving the knowledge of the orbit and securing the connection with the Phoenix part of the stream, kinematic data will be able to clarify 
the existence, or not, of sub-halo induced gaps in the stream. 

\section{The Phoenix-Hermus Stream}

\citet{GrillmairCarlberg:16} proposed that the Hermus stream  \citep{Grillmair:14} and the Phoenix 
stream  \citep{Balbinot:16} are quite plausibly parts of a single long stream. 
Although \citet{Sesar:15} proposed that the Phoenix stream is a recently 
disrupted cluster and \citet{Li:16} proposed an association with a more diffuse structure in Eridanus, 
the attraction of the Phoenix-Hermus association is that it is a single simple system which perhaps
surprisingly accurately combines two completely independent sets of sky position and velocity data. 
The two stream segments are both metal poor,  helping to bolter the confidence  of single common progenitor. 
However the available photometric measures are relatively low signal to noise and the moderate resolution spectra
do not give high precision velocities so that more and better quality data will strengthen (or not) the association.
Like most thin streams, the old stellar population and low metallicity is consistent with the 
progenitor system being a globular
cluster which would create a two armed stream 
that would wrap approximately half way around the galaxy.  That is, considering Phoenix and Hermus 
as single stream implies a fairly conventional origin.

The derived orbital eccentricity of the combined stream is $e\simeq 0.05$ \citep{GrillmairCarlberg:16} 
which means that there is relatively low
angular momentum spread amongst the stars pulled from the progenitor cluster (not visible, or, now dissolved) into 
the stream, hence the differential velocities  across the stream are 
relatively low so that the gaps created in sub-halo encounters will persist
for many more orbits than in a high eccentricity stream \citep{Carlberg:15a} making Phoenix-Hermus
a relatively attractive target for additional observational studies. At a galactocentric distance of 20 kpc and an inclination
of 60\degr\ the orbit does not encounter any significant molecular clouds \citep{Rice:16}. This paper measures the 
density variations in the Hermus section of the stream and explores the expected velocity variations associated
with sub-halos. The key assumption is that stream segments originate from a progenitor on a low eccentricity orbit, 
which the Phoenix-Hermus association helps motivate, but is not a requirement.

\begin{figure}
\begin{center}
\includegraphics[angle=-90,scale=0.7]{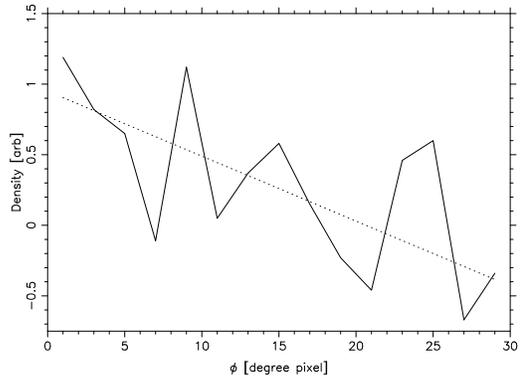}
\end{center}
\caption{The background subtracted stellar density in the Hermus segment 
of the Phoenix-Hermus stream, measured from the image of \citet{Grillmair:14}.
The error at each point is 0.46 units. 
The line shows the best linear fit, whose slope is significant at $3.3\sigma$.
The remaining density variations are not statistically significant.
}
\label{fig_denfit}
\end{figure}

The sky density of the Hermus section of the stream  visibly declines towards the south along its length.
The stream density measured in the northern piece of the Hermus stream as identified in the map of 
\citet{Grillmair:14}  is shown in Figure~\ref{fig_denfit}. There is a separate piece of Hermus
to the south whose association with Hermus is discussed as tentative
in \citet{Grillmair:14} and we do not include in our sky density measurement.
The density error at
each point in the 2\degr\ binning of Figure~\ref{fig_denfit} is 0.46 units as estimated from the local
background variations. 
A linear fit to these data finds the density as a function of angular 
distance to be 
$0.95\pm0.23 - (0.046\pm0.014)\phi$ where $\phi$ is measured in degrees and the density scale
is arbitrary. The decline from the largest value at the beginning
of the stream (at $\eta$= 36.05, $\lambda$= 40.85) to zero occurs over 20\degr, or 0.36 radians. 
Neither Pal~5 (beyond the initial 6\degr\ and over the 20\degr\ where the stream is readily visible) nor GD-1 have a significant mean density decline along their length \citep{Carlberg:12,Carlberg:13}, so
the finding that Hermus has a  density decline along the stream,  significant at the $3.3\sigma$ level,
is somewhat novel in this small group of three well-studied, thin streams. 
Once the slope is removed
there is no remaining excess variance in the stream density, although the noise levels are very high.
The decline in the density could either
 be the result of the expected changes of density along the orbit, or, could be a sub-halo induced gap, 
 which we will explore within our simulations.
The photometric data will improve 
as additional imaging comes available and {\it GAIA} astrometry will be able to identify brighter members of the stream. 
However, the key to really understand stream density variations is to obtain velocities, for which we explore
the expectations in the remainder of this paper.

\section{Phoenix-Hermus Stream Simulations }

\begin{figure}
\begin{center}
\includegraphics[angle=-90,scale=0.7]{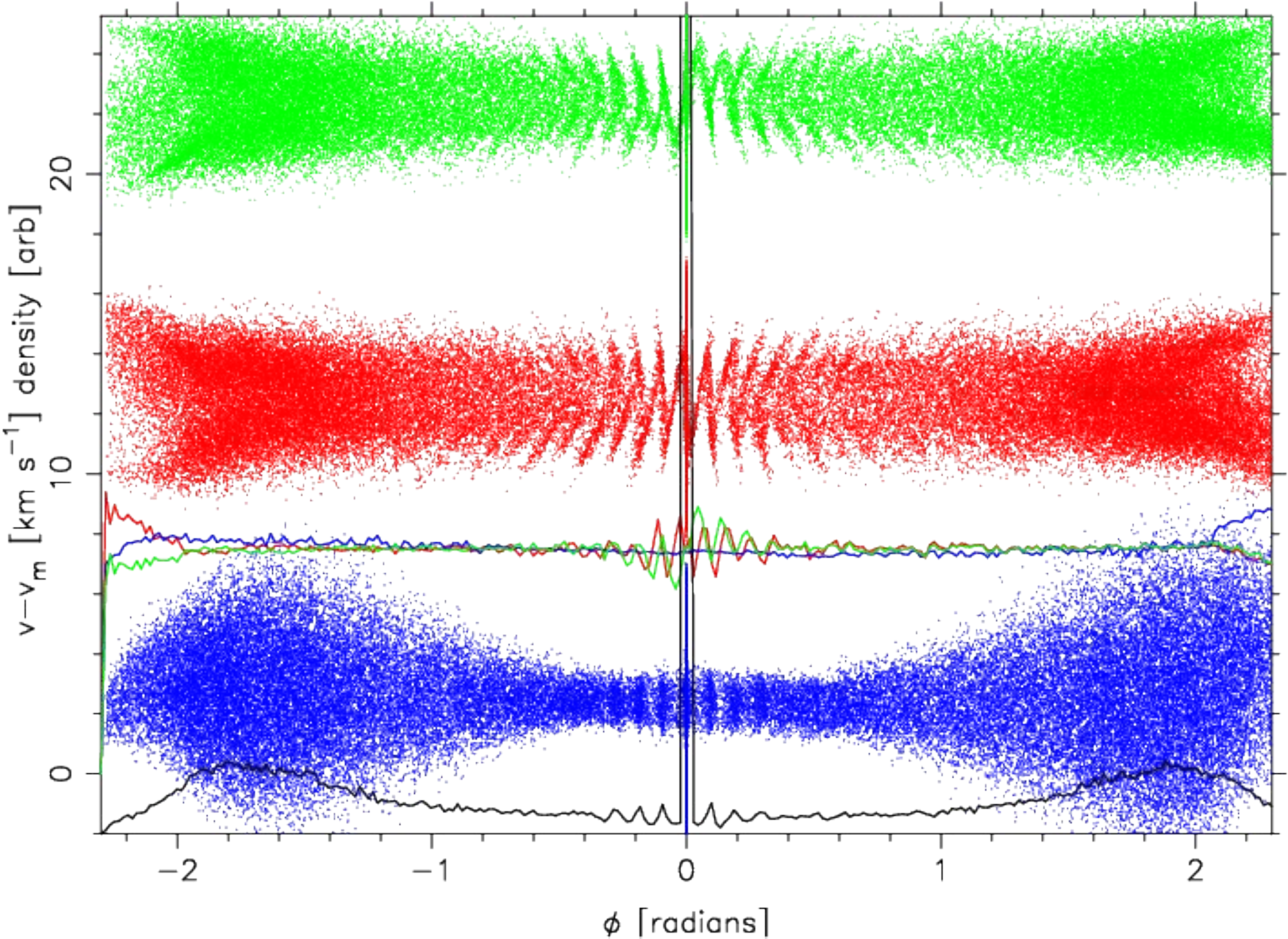}
\end{center}
\caption{The angular density distribution (black line at the bottom) and the velocities of a simulation 
with no sub-halos, as seen from the center of the potential.
The velocities are measured relative to the large scale mean values of  the stream at that angle.
The points show individual particles (blue, bottom, perpendicular to orbital plane; red, middle, 
radial velocity, and green, top, parallel to the orbital plane)
and the lines show the means in angular bins of 1\degr, with a running mean over 20\degr\ subtracted.
}
\label{fig_vxyz_0}
\end{figure}

A set of n-body simulations to create streams representative of the Phoenix-Hermus stream are done using the techniques 
described in \citet{Carlberg:16}.  The models are not matched to 
either the orbital or sky density data but are intended to explore the range of outcomes
in sub-halo filled potentials would give.
The model places a King model globular
cluster in the the MW2014 potential \citep{Bovy:15} at a location of x = 10 kpc, y = 0, and z= 17.32 kpc, 
that is,  a
galactocentric radius of 19.5 kpc and an inclination of 60\degr\ which is at the observationally
determined stream apocenter. No progenitor globular
cluster has been identified so the angular phase of the stream is essentially unconstrained. 
The model
cluster is started with a purely tangential velocity (in the y direction) of 178 \kms\ which gives rise to 
an orbit with an eccentricity close to the inferred value of 0.05.  The King model cluster has a mass
of $0.9\times 10^5 \msun$ which is about the lowest mass cluster that will give tidal streams extended enough
in angle, which scales as $M^{1/3}$, to provide substantial stream densities in opposite directions on the sky 
as in the Phoenix-Hermus association. The
stream width also scales as $M^{1/3}$, consistent with the 220 pc width that \citet{Grillmair:14} finds for Hermus, which
is relatively wide for a globular cluster stream.
The model cluster starts with an outer radius of 0.112 kpc which is equal to the nominal tidal radius estimated
from the local gravitational acceleration. Exactly filling
the tidal radius gives a fairly steady mass loss which totals a modest 11\% over an orbital lifetime of 10.7 Gyr. 
A modest over-filling of the tidal radius will boost the mass loss rate, but those results are not reported here. 
The simulation uses a shell code \citep{Carlberg:15b} which gives the same accuracy as a full n-body code.
500,000 particles are in the initial cluster.

The velocities and densities in the star stream with no perturbing
dark matter sub-halos are shown in Figure~\ref{fig_vxyz_0}. 
The density and mean velocities along the stream are measured in 1\degr\ bins, where angles are measured
relative to the satellite location along a great circle defined by the instantaneous angular
momentum vector of the satellite.  The velocities are first measured relative to the
values that the satellite orbit has at that location.
The particles emerge from the inner and outer saddle points in the potential and stream away, offset from
the orbit of the satellite. To allow for the offset and other large scale orbital variations
we smooth the mean particle velocities with a running average over $\pm$20\degr\
which is subtracted from the individual particles to give the offset from the running mean. 
The running mean becomes noisy
at the ends of the stream. 
Using a smaller window to compute the running mean reduces the end problems, but
also artificially suppresses the velocities offsets that sub-halos create.
Figure~\ref{fig_vxyz_0} shows significant epicyclic orbit 
density variations for about four cycles away from the progenitor in both directions, quickly decreasing with 
distance from the progenitor as differential
velocities in the stream cause them to be overlapped on the sky.

\section{The Effects of Sub-halos}

Sub-halos are added to the MW2014 potential following the prescription of \citet{Carlberg:16}. 
Only the inner 100 kpc of the simulation is filled
with sub-halos, with the inner 56 kpc fully populated
since there is no value in computing the forces from sub-halos that will never be near the path of the stream. 
Sub-halos are generated in the mass range $2.7\times 10^8 \msun$ (above which mass sub-halos are very rare) to
$2.7\times 10^5 \msun$ (at which mass sub-halos are no longer doing anything significant to a stream). The outcome is 
a total of about 100 sub-halos in a simulation matched to the Aquarius \citep{Aquarius} sub-halo distribution. 
The numbers of sub-halos in 40 realizations range from 57 to 168 for the different realizations from the same distribution.
We generate sub-halos with the constraint that the total mass is the expected mass in the selected part of the mass spectrum. 
This approach does not include an allowance for differences in the sub-halo populations in cosmological dark halos. 
\citet{Ishiyama:09} find that the fractional variation from one halo to another is about 25\% and \citet{Chen:11}
 find that inside about a tenth of the virial radius the variation from 
halo to halo is about a factor of 2, up and down relative to the mean. 
 If there were twice as many sub-halos in the region of the simulated stream, then
the number of significantly disturbed regions of the stream would exactly double.

\begin{figure}
\begin{center}
\includegraphics[angle=-90,scale=1.1]{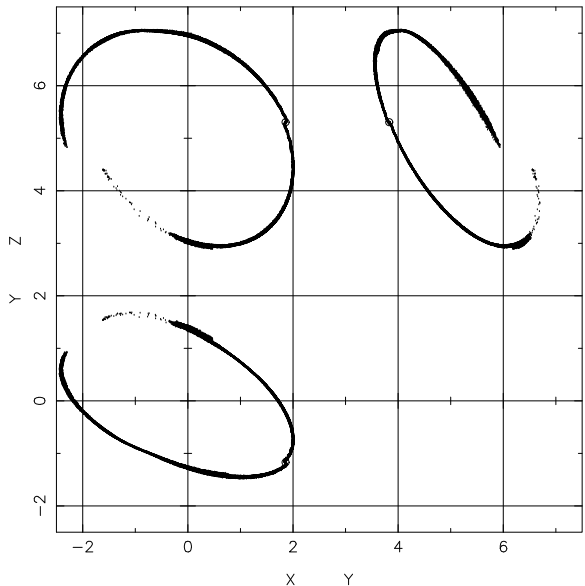}
\end{center}
\caption{The final moment of a simulation with sub-halos projected into galactic coordinates. The progenitor is
marked with a circle. A massive sub-halo passed through the stream region near x=0, y=1.5, z=2, where one unit
is 8 kpc. The upper and right plots have centers offset 5 units.}
\label{fig_xyz}
\end{figure}

Figure~\ref{fig_xyz} shows one of the streams projected onto galactic Cartesian coordinates at the final step of the simulation.
Because the progenitor orbit is nearly circular the stream properties
do not depend much on the orbital phase of the cluster. The particular  moment
shown in Figure~\ref{fig_xyz} 
is not a good match to the present time for  the Phoenix-Hermus stream since the progenitor cluster 
would be readily visible in the northern sky. 
The stream shows has had an encounter with a massive sub-halo near the end of the trailing stream a few  orbits earlier 
(the progenitor is rotating counter-clockwise) which creates a folded section of stream and a thin trailing segment.

\begin{figure}
\begin{center}
\includegraphics[angle=0,scale=1.0]{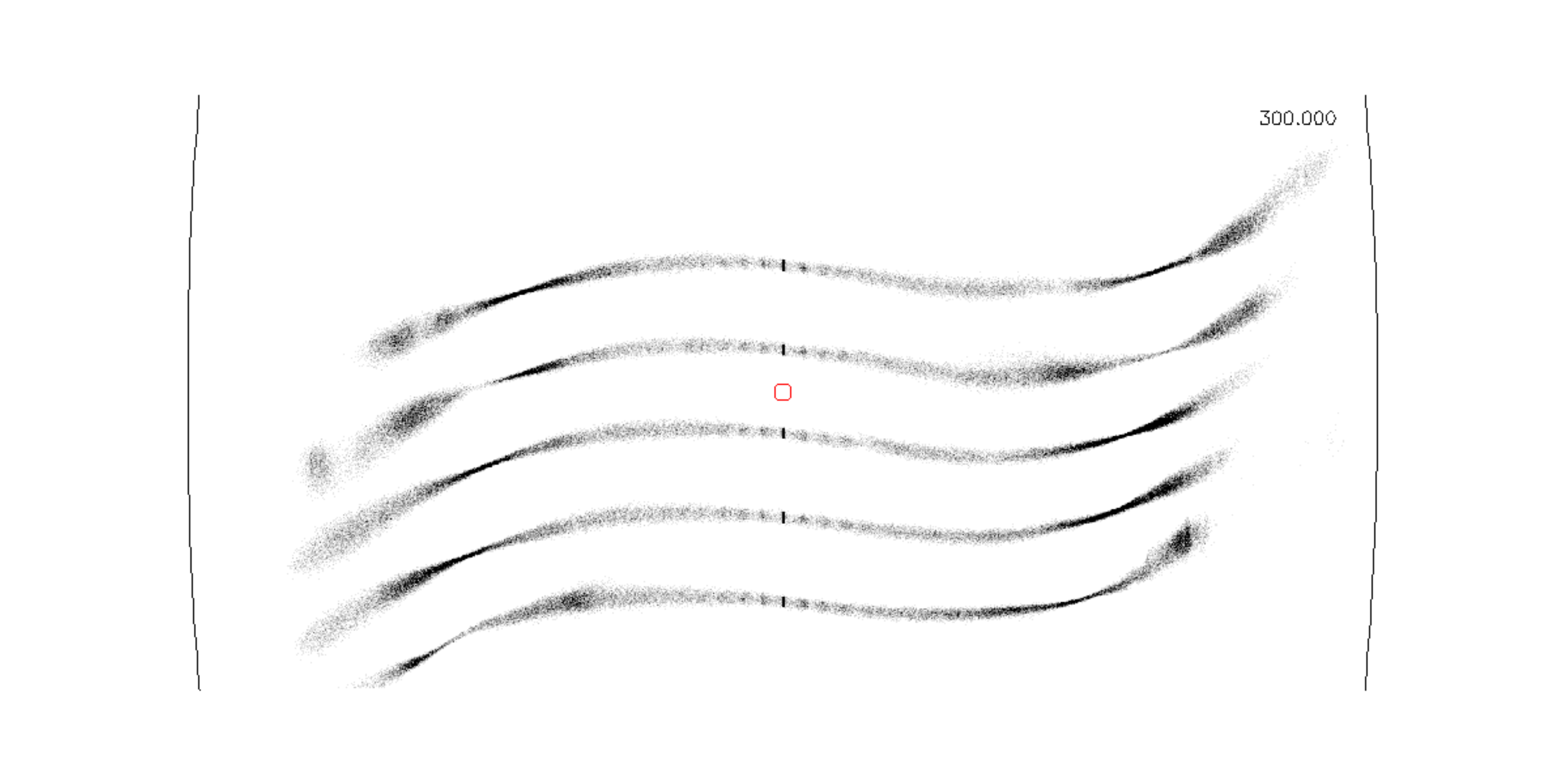}
\end{center}
\caption{Five streams plotted along their instantaneous orbital plane as projected onto the sky in an equal
area projection. 
The plots are offset 0.35 radian and extend from
over $\pm \pi$. The vertical scale is multiplied by a factor of five to make the width structure visible.
The lines at the side are the $\pm\pi$ edges of the coordinate system.}
\label{fig_tp}
\end{figure}

Figure~\ref{fig_tp} shows the projection of five separate realizations of the stream projected onto the sky.
The angles are measured with respect to the instantaneous orbital plane as defined by the total angular momentum vector 
(which is not a conserved quantity in the MW2014 potential) 
centered on the location of the progenitor cluster. The plots are offset from
one another 0.35 radian. The vertical displacements from the orbital plane of individual particles 
are multiplied by 5 to increase the visible
width of the streams. Near the progenitor the stream particles are young and the epicyclic oscillations of tidal loss dominate.
Further down the stream differential orbital velocities
 across the stream has blurred out the epicyclic features and the streams are old enough
that they have suffered a few encounters with sub-halos. The streams vary in width along their length, although
this is hard to detect in the presence of substantial foreground. Each simulation has at least one readily visible 
``gap'' in its longitudinal density.

\begin{figure}
\begin{center}
\includegraphics[angle=-90,scale=0.8]{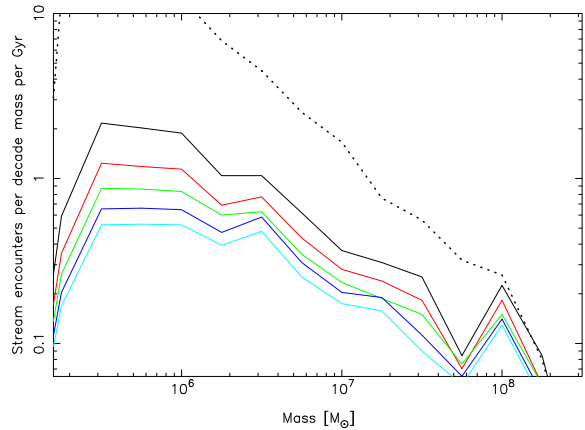}
\end{center}
\caption{Rate of stream hits over a 3~Gyr interval  starting at 7~Gyr. The
rate is shown out to a distance of 1, 1.5, 2, 2.5 and 3 times the scale
radius (solid lines from the bottom to the top) of the sub-halo at that mass.
The distribution of sub-halo masses is shown as the dotted line. The counting procedure
normalizes the distance to the sub-halo scale radius, which leads to the dip near the largest mass bin.}
\label{fig_nstream}
\end{figure}

Figure~\ref{fig_nstream} shows the distribution of sub-halo stream encounters at distances 
from 1 to 3 times
the scale radius of the sub-halo, the latter being a somewhat generous distance to produce a significant gap  \citep{Carlberg:12}. 
The procedure finds the sub-halos closest to a  stream particle, relative to its scale radius, at
each step.  The adopted counting procedure means that lower mass sub-halos are usually
not counted when a massive sub-halo is close to the same part of the stream. 
Figure~\ref{fig_nstream} has 4 mass bins per decade of mass.
The most massive sub-halos will typically have one stream hit in the last 3 Gyr during which time the
gaps fully develop and remain readily visible.

\begin{figure}
\begin{center}
\includegraphics[angle=-90,scale=0.7]{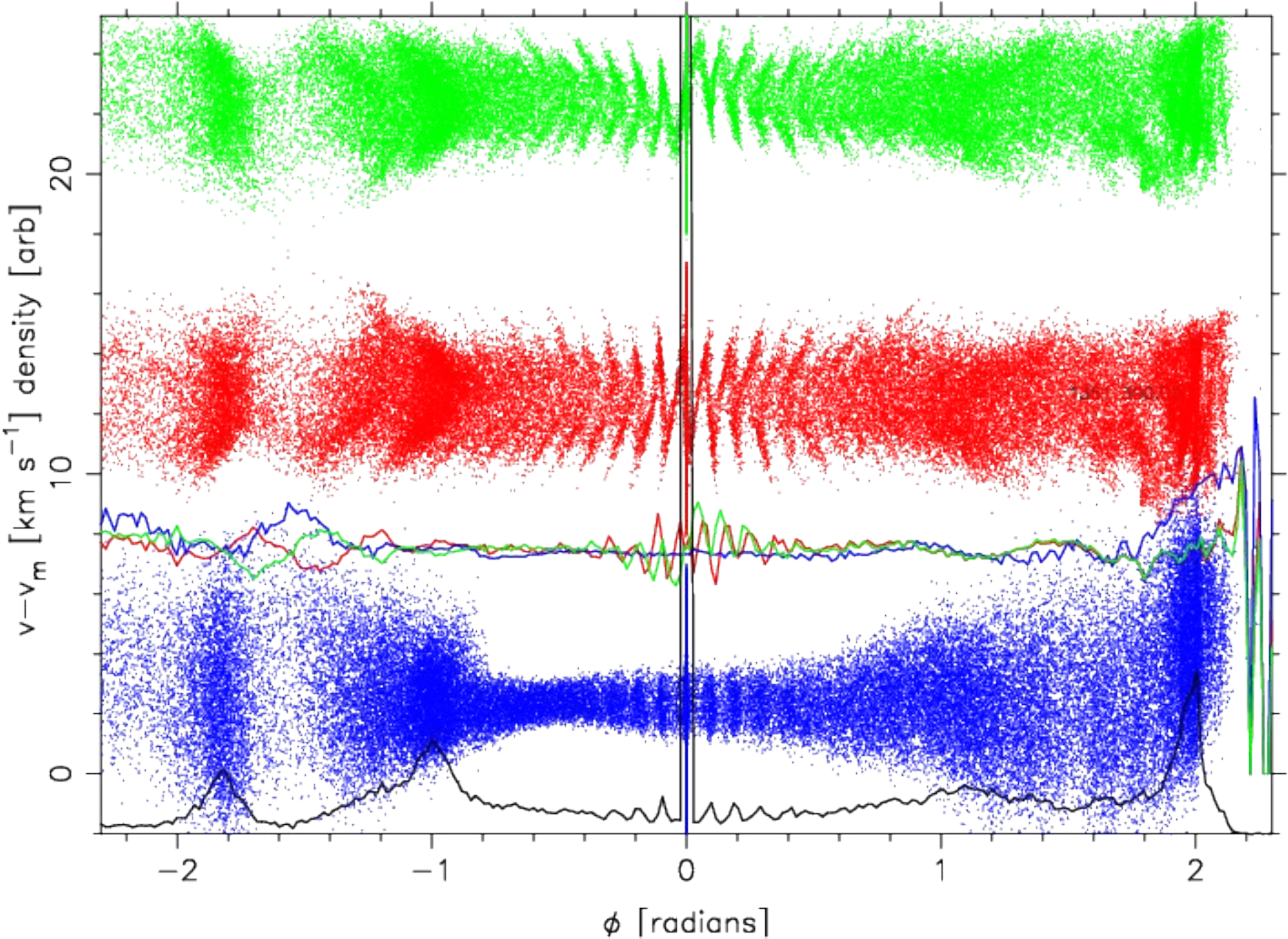}
\end{center}
\caption{The same density and velocities plotted as Figure~\ref{fig_vxyz_0} but 
for a simulation with sub-halos included. Even though
the individual particle velocities are scattered on the right hand side (trailing) part of the
stream, the
mean velocities in projection do not show as large a deviation as the left hand side.}
\label{fig_vxyz_sub}
\end{figure}

\section{Sub-halos and Stream velocity perturbations }

Figure~\ref{fig_vxyz_sub} shows the same density and velocity measurements as Figure~\ref{fig_vxyz_0} 
but now for a simulation
with sub-halos. Comparing the two shows the effects of the sub-halos on the stream 
are readily visible beyond about half the stream length, since that is where the stream
is old enough to be likely to have had an encounter with one of the heavier sub-halos. 
The mean radial, parallel and perpendicular velocities are shown as the lines. 
Radial velocity measurements 
have no sensitivity to distance errors and are the most likely to be practical in the immediate future. 
For completeness we also show the local tangential and vertical velocities.
If we had
access to high precision velocities for many individual stars in the stream the presence (or not)
of sub-halos would be obvious. The mean velocities show much smaller effects, with 
offsets of about 1 \kms\ around the locations of the large gaps. Only the largest sub-halos, those
making gaps 5-10\degr\ in size, are readily visible. Small gaps simply do not show up as visible,
a consequence of sky overlap of gapped and non-gapped stream regions, and
the speed which differential orbital motion, even in this fairly circular orbit, blurs out small gaps.

\begin{figure}
\begin{center}
\includegraphics[angle=0,scale=1.1]{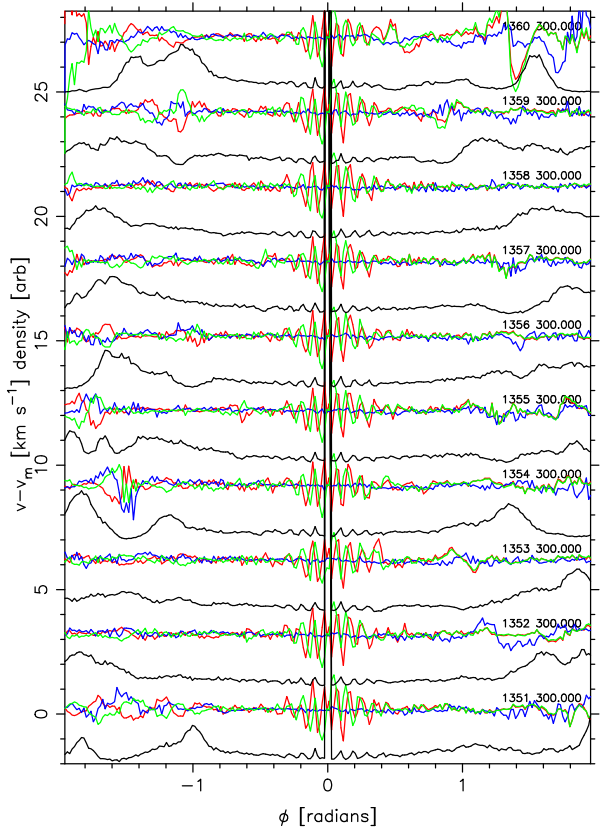}
\end{center}
\caption{The same as Figure~\ref{fig_vxyz_sub} but with 9 additional realizations of the stream with sub-halos.
The black line is the density field and the red, blue and green lines are the radial, local vertical and local tangential
velocities, respectively.
The figure illustrates the range of mean velocity amplitudes and locations of offsets relative to density variations
in the stream.}
\label{fig_vxyz}
\end{figure}

Figure~\ref{fig_vxyz} shows ten different realizations of the same standard sub-halo population to 
illustrate the range of mean velocity offsets that are exhibited for the same progenitor cluster orbit and
different stastical realizations of the sub-halo population. In general if there is a feature
that looks like a density gap, with excess density on either side, there is a good chance that there
will be velocity offsets associated with it. However, there are exceptions from time to time. This
means that a single stream may provide strong evidence of sub-halo encounter, but it will be important
to have velocity measurements for as many streams as possible, to allow for case to case differences.
Velocities in the plane of the sky perpendicular and along the stream are affected by distance errors, 
but the means remain unaffected. 
The stars along the stream contain interesting information about the sub-halo mass, velocity and time of 
encounter \citep{Erkal:15b}.

Given the radial velocity signature of a sub-halo encounter, 
the observational requirement is that to measure reliably an offset from the mean velocity of 1~\kms, spread
over 10\degr, with velocities having a peak-to-peak spread of 10~\kms\ will require individual 
stellar velocities to be 0.5 \kms\ or better, for approximately 10 stream stars per degree length of stream. 
More stars and better velocities will begin to identify the internal structure of the stream, which provides
significant additional information on the sub-halo interaction. Detailed modeling for
specific observational and stream orbit parameters is straightforward using models similar to those presented here.

The straight line fit to density measurements along the Hermus section of the stream
found a decline from the first reliable points to zero over 20\degr\ angular extent of stream. Such a steep decline is not
seen in the no sub-halo simulation, Figure~\ref{fig_vxyz_0}.  The ends of the stream have the steepest 
density decline, but the ends are very artificial, being the result of instantaneously inserting a 
progenitor star cluster into a Milky Way orbit, rather than some more gradual introduction through, 
say, a merger process. On the other hand,
density declines as steep as measured and more are frequently obtained in simulations with sub-halos,
see Figure~\ref{fig_vxyz_sub}.

\section{Discussion}

The Phoenix-Hermus stream, whether it is a single object or not, is a good case to consider for sub-halo interactions
because the two visible 
stream segments are long, thin, and appear to be on a relatively circular orbit 
allowing dark matter sub-halo induced density
variations, and associated velocity variations,
 to remain visible for many orbits before differential orbital motions blur them out.
The main point established here is that we expect 
about one gap per stream to be sufficiently large,  about 10\degr,
that there will usually be mean radial velocity offsets of 1-2 \kms.   Smaller size gaps 
are present but would require many high precision velocities and dynamical modeling to interpret.
The velocity deviation profiles along the stream
 do not have any simple universal shape and although
associated with a density variation, the precise location along the stream relative 
to a density gap varies fairly uniformly from one edge of a gap to the other over the typical 10\degr\ range.

For the expected mean sub-halo population there will be only one easily
visible gap present at a time and that gap is the outcome of an encounter of the stream 
with one of the larger sub-halos, $\approx 10^8 \msun$. 
Lower mass sub-halos, those below about $10^7 \msun$, generally create gaps and velocity changes that are too small to persist.  With only one arm of a stream typically visible, it is possible that no detectable sub-halo effects 
will be found. However, in the inner halo there is a substantial range for the
predicted sub-halo abundance. With the LMC approaching for the first time \citep{Besla:07} it is possible that its collection of outer sub-halos is currently substantially boosting the numbers of sub-halos in the Milky Way as they stream through on the first infall. 
For Phoenix-Hermus the ideal is to have a velocity survey everywhere along the stream, but the best locations for finding velocity changes is where the density is changing significantly.

\acknowledgements

This research was supported by CIFAR and NSERC Canada. An anonymous referee provided constructive criticism.

\end{document}